\documentclass[a4paper]{jpconf}
\usepackage{graphicx}
\usepackage{amsfonts}
\usepackage{amsmath}
\usepackage{bm}
\newcommand{\be}{\begin{equation}}
\newcommand{\ee}{\end{equation}}
\newcommand{\bea}{\begin{eqnarray}}
\newcommand{\eea}{\end{eqnarray}}
\newcommand{\hf}{\frac12}
\newcommand{\nn}{\nonumber\\}
\def\eq#1{(\ref{#1})}
\def\journal#1#2#3#4{#4 {\em #1} {\bf #2} #3}

\def\fd#1#2{\frac{\delta#1}{\delta#2}}

\def\la{\langle}
\def\ra{\rangle}

\def\mr#1{{\mathrm{#1}}}
\def\ord#1{{\cal O}(#1)}

\def\bre{\hskip-4pt/}
\def\dv#1{\dot{\v{#1}}}
\def\ddv#1{\ddot{\v{#1}}}

\def\ha{{\hat a}}
\def\hj{{\hat j}}

\def\hpsi{{\hat\psi}}
\def\hsib{\hat{\bar\psi}}
\def\hpsid{{\hat\psi^\dagger}}

\def\hx{\hat x}
\def\hy{\hat y}
\def\hD{{\hat D}}
\def\hG{{\hat G}}
\def\hJ{{\hat J}}
\def\ih{\frac{i}{\hbar}}

\def\psid{\psi^\dagger}
\def\sign{\mr{sign}}

\def\v#1{{\bm{#1}}}

\def\Tr{{\mr{Tr}}}

\begin{document}
\title{Irreversibility and decoherence in an ideal gas}

\author{Janos Polonyi}

\address{Strasbourg University, CNRS-IPHC, 23 rue du Loess, BP28 67037 Strasbourg Cedex 2, France}

\ead{polonyi@iphc.cnrs.fr}

\begin{abstract}
Different models are described where non-interacting particles generate dissipative effective forces by the mixing of infinitely many soft normal modes. The effective action is calculated for these models within the Closed Time Path formalism. This is a well known scheme for quantum systems but its application in classical mechanics presents a new, more unified derivation and treatment of dissipative forces within classical and quantum physics.
\end{abstract}

\section{Introduction}
The derivation of dissipative forces is a challenge for us on two counts. First, these forces are not covered by the fundamental equations of motion. Dissipation, a special kind of breakdown of the time reversal invariance, arises from the restriction of our attention to a subset of a closed system, in short it is an effective force. Second, the usual formalism of classical or quantum mechanics, based on Lagrangians and Hamiltonians, can not cope with such non-conservative forces. 

Some attempts to derive friction dissipative forces and decoherence are surveyed here by calculating the effective action for the observed system within the Closed Time Path (CTP) formalism, well known in quantum field theory \cite{schw,mahanthappa,bakshi,keldysh}. Its extension for classical mechanics \cite{arrow,galley,effth,galleytsang}, which provides a unified treatment of effective forces on the classical and the quantum level and offers a fresh look into the quantization of open systems, is surveyed in the first part of this paper. The second half contains a brief discussion of dissipative effective forces in two classical models, in a set of harmonic oscillators and in classical electrodynamics, followed by the quantum mechanical examples, a test particle in an ideal gas and the dynamics of the probability current in an ideal fermi gas.

\section{Dissipative forces in effective theories}
Let us consider a classical model, described by the system and the environment coordinates $x$ and $y_1,\ldots,y_N$, respectively. The closed dynamics, defined by the action $S[x,y]$, is supposed to be symmetric under the inversion of and the translation in the time. The solution of the equations of motion, 
\be\label{oeomse}
\fd{S[x,y]}{x}=0,~~~~~~\fd{S[x,y]}{y}=0,
\ee
are made unique by some auxiliary condition, to be chosen as initial conditions. To find the effective system equation of motion one solves the second equation for a general system trajectory and inserts the solution, $y[x]$, into the first equation. One can gain more insight into the effective dynamics by calculating the effective action, $S_{eff}[x]=S[x,y[x]]$,  whose variational equation,
\be\label{veeom}
\fd{S_{eff}[x]}{x}=0,
\ee
is the effective equation of motion. 

But this procedure is formal an incomplete. The problem comes from the way the auxiliary conditions are handled in the variational method: The Euler-Lagrange equation is derived by fixing the initial and final coordinates but we have initial conditions in this problem. This is an important issue for dissipative effective dynamics where the final environment coordinates are extremely complicated. An extension of the traditional variation method is needed at this point. The proposition, put forward in this Section, is to let ourself guided in establishing an extension of the classical variational method by the already well known solution of this problem in quantum theory, namely by the use of mixed states. This strategy not only lead us to an action principle which covers dissipative forces but also provides a fresh view of the standard quantization rules when applied for open systems.

\subsection{Semiholonomic force}\label{semihols}
One of the main obstacle to describe effective dynamics is that it contains non-conservative forces. Analytic mechanics starts with d'Alembert principle, stating that the virtual work of the external and the inertial forces is always vanishing along the trajectory, 
\be
(F_{ext}-m\ddot x)\delta x=0.
\ee
This equation can be converted into a much more powerful form by assuming that the force is holonomic, namely it can be obtained from a scalar function, $U(x,\dot x)$ by derivation, 
\be\label{hol}
F(x,\dot x)\delta x=-\delta x\partial_xU(x,\dot x)-\delta\dot x\partial_{\dot x}U(x,\dot x).
\ee
Hamilton considered the time integral of d'Alambert principle for a holonomic force and noticed that it can be written as a variational equation,
\be
0=\delta\int_{t_i}^{t_f}dt\left[\frac{m}2\dot x^2-U(x,\dot x)\right]-\delta x(m\dot x+\partial_{\dot x}U)\biggr|_{t_i}^{t_f}.
\ee
Note that the initial and final variations cancel in this equation. To underline the formal similarities of the classical and quantum dynamics we continue with the usual variational principle,
\be\label{usualvp}
0=\delta\int_{t_i}^{t_f}dt\left[\frac{m}2\dot x^2-U(x,\dot x)\right].
\ee

The holonomic forces are conservative and we need a generalization of \eq{hol} for effective  forces. The coordinate and velocity dependence dependence of $U(x,\dot x)$ in \eq{hol} provides us an expression of the force and assures the energy conservation. We separate these two roles by introducing passive and active copies of the degrees of freedom,
\be\label{doublers}
x\to\hx=(x^+,x^-),
\ee
the active coordinate being used to calculate the force and the passive coordinate controlling the potential in an independent manner. The result is expression,
\be
F(x,\dot x)\delta x=-[\delta x\partial_{x^+}U(\hx,\dot{\hx})+\delta\dot x\partial_{\dot x^+}U(\hx,\dot{\hx})]_{|x^+=x^-=x},
\ee
for the semiholonomic forces. These forces are non-conservative and it will be argued below that they cover all effective forces, arising within a subsystem of a closed system.

\subsection{Initial conditions}
Another complication of the effective dynamics is related to the missing auxiliary conditions. Let us suppose that the environment is finite, $N<\infty$, when the effective equation of motion,
\be\label{effeomho}
\left[c_0+c_2\partial_t^2+\cdots+c_{2(N+1)}\partial_t^{2(N+1)}\right]x=0.
\ee
can easily be derived in harmonic models by the use of normal modes. The general solution of this equation requires the knowledge of $2(N+1)$ auxiliary conditions of which two are known, $x(t_i)$ and $\dot x(t_i)$, the environment being unobserved. How to solve this equation with the insufficient number of auxiliary conditions at hand? It is easy to see that this problem is not restricted to harmonic models. 

The missing initial conditions may lead to yet another, more particular problem of the effective equation of motion \eq{effeomho}. This equation is formally invariant under time reversal but this symmetry is bound to be broken in the effective theory. In fact, to check the the time reversal invariance of the effective theory one records the motion of the system and checks whether the motion, seen by playing the record backward in time, satisfies the same effective equation of motion. The point is that the environment initial conditions influence the effective equation of motion in an implicit manner and the initial environment conditions of the time reversed record are the final environment conditions of the original motion. 

We propose the use of variational principle as a remedy of these problems. More precisely, the solution of the effective equation of motion will be constructed by the help of the retarded Green function, derived from the effective action.

\subsection{Classical chrono-dynamics}
\begin{figure}[h]
\includegraphics[width=20pc]{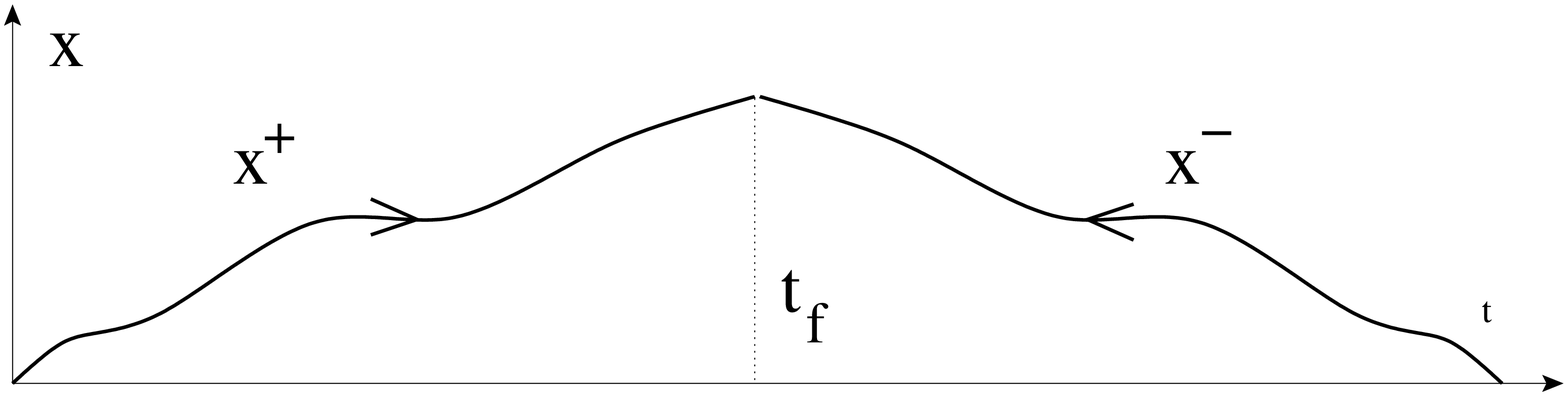}\hspace{2pc}%
\begin{minipage}[b]{14pc}\caption{\label{ctppath}The system undergoes a time reversal transformation, the time arrow is flipped at $t=t_f$, and the motion is followed back to the initial state.}
\end{minipage}
\end{figure}

To avoid the need of fixing the final coordinates and to allow the dynamical breakdown of time reversal invariance we follow the motion in an extended time interval. We start the motion at $t_i$ with some initial conditions, perform a time reversal at $t_f$ and follow our system back in time until it reaches its initial state, c.f. Fig. \ref{ctppath}. The result is a closed time path, $\tilde x(t)=x(t)$, for $t_i<t<t_f$ and $\tilde x(t)=x(2t_f-t)$ when $t_f<t<2t_f-t_i$. This notation is rather cumbersome and a simpler book-keeping is provided by the introduction of the CTP doublers, \eq{doublers}, with trajectory $\hx(t)=(x^+(t),x^-(t))=(\tilde x(t),\tilde x(2t_f-t))$. This step sheds more light on the role of the active and passive copies, introduced above: The non-conservative part of the effective force arises from the new segment of the extended motion and the active and the passive coordinates belong to the same degrees of freedom because the equation $x^+(t)=x^-(t)$ holds after having imposed the equation of motion. Note that the time arrow is oriented in the opposite way for the two trajectory. This is the main achievement of this formalism, the possibility of following the motion in both directions of the time, and is the source of the name chronon, introduced for the doublet $\hx$.

Both member of the chronon, $x^\pm$, are dynamical variables therefore one needs an action, $S[\hx]$, to generate their dynamics. The obvious choice for a closed system is 
\be\label{naivea}
S[\hx]=\int_{t_i}^{t_f}dt[L(x^+(t),\dot x^+(t))-L(x^-(t),\dot x^-(t))],
\ee
where $L(x,\dot x)$ denotes the traditional Lagrangian. The minus sign arises from the opposite time arrow of the trajectories $x^+(t)$ and $x^-(t)$ and introduces a symplectic structure for the CTP index. The same initial conditions are imposed for both members of the chronon. The boundary term in the variational equation of motion, \eq{usualvp},  should not be canceled in the usual way, by requiring $\delta x^\pm(t_f)=0$, rather by imposing the constraint
\be\label{ctpconstr}
x^+(t_f)=x^-(t_f),
\ee
which leaves the final state of the motion free, an essential requirement for the environment of dissipative systems.

There is a degeneracy in the action at $x^+(t)=x^-(t)$ which has to be split in order to arrive at unique Green functions. For this end we add an infinitesimal imaginary part to the CTP Lagrangian,
\be
L(\hx,\dot{\hx})=L(x^+,\dot x^+)-L(x^-,\dot x^-)+L_{spl}(\hx,\dot{\hx}),
\ee
with $L_{spl}(\hx,\dot{\hx})=i\epsilon(x^{+2}+x^{-2})/2$. The advantage of removing the degeneracy with an imaginary term is the CTP symmetry, the transformation
\be\label{ctpsym}
S[\tau\hx]=-S^*[\hx]
\ee
of the action under the exchange, $\tau(x^+,x^-)=(x^-,x^+)$, which flips the time arrow.

\subsection{Green functions}
Let us first consider the Green function of a harmonic system, given by the action  $S[\hx]=\hx\hat K\hx/2+\hx\hj$ where $\hj=(j^+,j^-)$ is an external source. We use a condensed notation where the trajectories are handled as vectors, the scalar product denotes the time integration and $\hx\hj=\sum_\sigma x^\sigma j^\sigma$. The CTP Green function is defined as $\hD=\hat K^{-1}$ and allows us to write the trajectory, corresponding to a physically realizable source, $j^+=-j^-$, as
\be
x(t)=-\sum_{\sigma'}\int dt'D_0^{\sigma\sigma'}(t,t')\sigma'j(t').
\ee
This expression is independent of $\sigma$ and leads to the CTP exchange symmetry, $D^{++}+D^{--}=D^{+-}+D^{-+}$, of the Green function. Similar relation is satisfied by $\hat K$, as well. These together yield the block structure
\be\label{blockg}
\hD=\begin{pmatrix}D^n&-D^f\\ D^f&-D^n\end{pmatrix}+iD^i\begin{pmatrix}1&1\cr1&1\end{pmatrix}
\ee
and
\be
\hat K=\hat\sigma\left[\begin{pmatrix}K^n&-K^f\cr K^f&-K^n\end{pmatrix}+iK^i\begin{pmatrix}1&1\cr1&1\end{pmatrix}\right]\hat\sigma,
\ee
where
\be
\hat\sigma=\begin{pmatrix}1&0\cr0&-1\end{pmatrix}
\ee
is the ``metric tensor'' of CTP. The application of these rules for the electromagnetic field suggests the definition $D^{\stackrel{r}{a}}\equiv D^n\pm D^f$ for the retarded and advanced Green functions. One can define these components for the inverse, $K^{\stackrel{r}{a}}\equiv K^n\pm K^f$, as well, and the relations $K^{\stackrel{r}{a}}=(D^{\stackrel{r}{a}})^{-1}$ and $K^i=-(D^a)^{-1}D^i(D^r)^{-1}$ follow. The symmetry, $K^{\sigma\sigma'}(t,t')=K^{\sigma'\sigma}(t',t)$, introduces similar equation for $\hD$ and the relations $D^n(t,t')=D^n(t',t)$, $D^i(t,t')=D^i(t',t)$ and $D^f(t,t')=-D^f(t',t)$.

The translation invariance in time is recovered by performing the limit $t_i\to-\infty$, $t_f\to\infty$ and the Green function of the harmonic oscillator, $L=m\dot x^2/2-m\Omega^2x^2/2$, turns out to be \cite{arrow}
\be
\hD(t,t')=\int\frac{d\omega}{2\pi}e^{-i\omega(t-t')}\hD_\omega(\Omega),
\ee
with
\be\label{doo}
\hat D_\omega(\Omega)=\frac1m\begin{pmatrix}\frac1{\omega^2-\Omega^2+i\epsilon}&-2\pi i\Theta(-\omega)\delta(\omega^2-\Omega^2)\cr
-2\pi i\Theta(\omega)\delta(\omega^2-\Omega^2)&-\frac1{\omega^2-\Omega^2-i\epsilon}\end{pmatrix}.
\ee
This result can be used to define the action in this limit, $K^n=m(\omega^2-\Omega^2)$, $K^f=i\sign(\omega)\epsilon$, $K^i=\epsilon$, $S[\hx]=S[x^+]-S[x^-]+S_{spl}[\hx]$, where the infinitesimal imaginary part,
\be
S_{spl}[\hx]=\frac{i\epsilon}2\int_{-\infty}^\infty dt[x^{+}(t)-x^{-}(t)]^2+\frac\epsilon\pi P\int_{-\infty}^\infty dtdt'\frac{x^+(t)x^-(t')}{t-t'},
\ee
handles the boundary conditions in time, and $P$ stands for the principal value prescription. Note that the coupling between the chronons at the final time, \eq{ctpconstr}, is replaced by infinitesimal, time translation invariant mixing terms.

Let us consider now a function of the coordinate, $A(x)$, and perform the Legendre transformation, $W[\hj]=S[\hx]+\hj\hat A(\hx)$, where $\delta\{S[\hx]+\hj\hat A(\hx)\}/\delta\hx=0$. The Green functions of $A(x)$ for an interacting system are defined as the coefficient functions of the expansion
\be
W[\hj]=\sum_{n=0}^\infty\frac1{n!}\sum_{\sigma_1,\ldots,\sigma_n}\int dt_1\cdots dt_nD^{\sigma_1,\ldots,\sigma_n}(t_1,\ldots,t_n)j^{\sigma_1}(t_1)\cdots j^{\sigma_n}(t_n),
\ee
and can be calculated by iteration. The inverse transformation, $S_{eff}[\hat A]=W[\hj]-\hj\hat A(\hx)$ and $\delta W[\hj]/\delta\hj=\hat A(\hx)$, gives the effective action. The solution of its equation of motion,
\be\label{solgrf}
\hat A(\hx(t))=\sum_{n=0}^\infty\frac1{n!}\sum_{\sigma,\sigma_1,\ldots,\sigma_n}\int dt_1\cdots dt_nD^{\sigma,\sigma_1,\ldots,\sigma_n}(t,t_1,\ldots,t_n)j^{\sigma_1}(t_1)\cdots j^{\sigma_n}(t_n),
\ee
is expressed by the help of the Green functions. Such a method to construct the solution of the effective equation of motion does not produce runaways trajectories as long as we are allowed to use the residuum theorem to calculate frequency integrals of the Green functions. This is how this scheme realizes Dirac's idea about the elimination of runaway solution for a point charge in classical electrodynamics \cite{dirac}.

\subsection{Effective action}
After having included the initial conditions into the variational principle we return to the effective action in the generic model, defined by the action $S[x,y]=S_s[x]+S_e[x,y]$. The steps, leading to eq. \eq{veeom} are now well defined and yield 
\bea
S_{eff}[\hx]&=&S_s[x^+]+S_e[x^+,y^+[\hx]]-S_s[x^-]-S_e[x^-,y^-[\hx]]+S_{spl}[\hx]+S_{spl}[\hy[\hx]]\nn
&=&S_s[x^+]-S_s[x^-]+S_{infl}[\hx]+S_{spl}[\hx],
\eea
where the trajectory $\hy[\hx]$ satisfies the equation of motion, $\delta S_e[\hx,\hy]/\delta\hy=0$. The second equation defines the influence functional \cite{feynman}, 
\be
S_{infl}[\hx]=S_e[x^+,y^+[\hx]]-S_e[x^-,y^-[\hx]].
\ee
A physically better motivated form of the effective action is
\be\label{ekchrc}
S_{eff}[\hat x]=S_1[x^+]-S_1[x^-]+S_2[\hx]+S_{spl}[\hx]
\ee
where the separation of the terms is rendered well defined by the condition $S_2[0,x^-]=S_2[x^+,0]=0$. The role of the different actions, introduced in this manner, becomes clear in the parametrization, $x^\pm=x\pm x^d/2$, whose advantage in classical mechanics is that it is sufficient to know the action in $S=\ord{x^d}$ since $x^+(t)=x^-(t)$ holds for the solution of the equations of motion. The variational equation of motion for $x^d$ at $x^d=0$,
\be
0=\fd{S_1[x]}{x}+\fd{S_2[x^+,x^-]}{x^+}_{|x^+=x^-=x},
\ee
represents the holonomic and the semiholonomic forces by the help of $S_1$ and $S_2$, respectively, and realizes the scheme of Section \ref{semihols}: The dynamics, generated by the single coordinate, $S_1$, is conservative and obeys Noether theorem and the coupling of the doubler, $S_2$, introduces the dissipative forces \cite{saso}. The chronon couplings are due to the influence of the final environment coordinate by the system trajectory and the CTP scheme reduces to two independent traditional action principles if the final environment coordinates are kept fixed during variation. The CTP symmetry, \eq{ctpsym}, is inherited by the effective action and provides the proof that the family of semiholonomic forces is closed with respect to generating effective interactions.

A distinguishing feature of the CTP formalism is that the system-environment interactions are mapped into the interactions within the chronon. This is an unexpected and a highly non-trivial result, holding for arbitrary large and complex environment. We shall see in the quantum case, discussed below, that the same chronon coupling encodes the system-environment entanglement, as well, making entanglement and semiholonomic forces identical.

\subsection{Quantum chrono-dynamics}
The CTP formalism was introduced by Schwinger to carry out the perturbation expansion for an expectation value $\bar A(t)=\la\psi(0)|U^\dagger(t)AU(t)|\psi(0)\ra$ \cite{schw}, where the need of the reduplication of the degrees of freedom arises from the simultaneous use of bra and ket, two virtually independent but physically identical state vectors. One is tempted to follow the naive quantization procedure of Schr\"odinger for the chronon and to introduce the chronon wave function, $\psi(x^+,x^-)$. This function is actually the density matrix, $\psi(x^+,x^-)=\rho(x^+,x^-)$, shedding a new light on Gleason theorem.

The more systematical, perturbative treatment is facilitated by the generating functional,
\be\label{ctpgf}
e^{\ih W[\hj]}=\Tr T[e^{-\ih\int dt(H(t)-j^+(t)x(t))}]\rho_iT^*[e^{\ih\int dt(H(t)+j^-(t)x(t))}],
\ee
where $\rho_i$ denotes the initial density matrix. Its path integral representation is
\be
e^{\ih W[\hj]}=\int D[\hat x]e^{\ih S[\hat x]+\ih\int dt\hj(t)\hat x(t)},
\ee
where the integration extends over chronon trajectories which satisfy the condition \eq{ctpconstr} and the action is the same as in the classical case. 

The effective dynamics can be found in a manner, similar to the classical case. The (Wilsonian) effective action is defined by
\be
e^{\ih S_{eff}[\hx]}=te^{\ih S_s[x^+]-\ih S_s[x^-]}\int D[\hy]e^{\ih S_e[x^+,y^+]-\ih S_e[x^-,y^-]+\ih S_{slp}[\hx]},
\ee
and the form of eq. \eq{ekchrc} applies again. 

One may consider a generalization where one leaves out the trace in the generator functional \eq{ctpgf}. The result is the Open Time Path scheme, an extension of the CTP formalism, based on the density matrix,
\be
\rho(x^+_f,x^-_f)=\la x^+_f|U(t_f,t_i)\rho_iU^\dagger(t_f,t_i)|x^-_f\ra
\ee
The path integral representation of the reduced density matrix,
\be
\rho(x^+_f,x^-_f)=\int D[\hx]D[\hy]e^{\ih S[x^+,y^+]-\ih S[x^-,y^-]+\ih S_{spl}[\hx]+\ih S_{spl}[\hy]},
\ee
where the trajectories satisfy the final conditions $\hx(t_f)=\hx_f$ and $y^+(t_f)=y^-(t_f)$. Noteworthy the effective action is the same as in the CTP scheme,
\be
\rho(x^+_f,x^-_f)=\int D[\hx]e^{\ih S_1[x^+]-\ih S^*_1[x^-]+\ih S_2[\hx]+\ih S_{spl}[\hx]}.
\ee
This form reveals that decoherence in the coordinate (field) diagonal representation consists of the suppression of the contributions to the path integral of well separated trajectories within a chronon and is generated by $\Im S_{infl}$. The couplings within a chronon, $S_2$, are due to the simultaneous contributions of several final environment states in the trace of \eq{ctpgf}, they render the system state mixed and encode the system-environment entanglement. An important role of the CTP formalism in quantum field theory is to visualize the elementary processes which contribute to the full reduced density matrix, rather than to the transition amplitudes between pure states, in terms of Feynman graphs.

\section{Effective models}
Now we look into model calculations where irreversibility, and decoherence in the quantum case, are generated by non-interacting modes. Irreversibility and decoherence will be detected by the dissipative terms in the effective equation of motion and the suppression of a chronon trajectory contribution to the path integral with well separated trajectories, respectively.

\subsection{Classical toy model}\label{toymods}
The simplest, non-trivial classical effective model corresponds to a set of linearly coupled harmonic oscillators, given by the Lagrangian \cite{caldeira},
\be\label{holagr}
L=\frac{m}2\dot x^2-\frac{m\omega^2_0}2x^2+\sum_{n=1}^N\left(\frac{m}2\dot y_n^2-\frac{m\omega^2_n}2y_n^2-g_ny_nx\right).
\ee
It is sometime more advantageous to write it in a form where the system action alone is already a good  approximation for the slow system dynamics, $\dot x\to0$,
\be\label{holagrc}
L=\frac{m}{2}\dot x^2-\left(\frac{m\omega_0^2}2-\sum_n\frac{g_n^2}{2m\omega^2_n}\right)x^2+\sum_n\left[\frac{m}2\left(\dot z_n-\frac{g_n}{m\omega_n^2}\dot x\right)^2-\frac{m\omega_n^2}2z_n^2\right].
\ee
The action will be written in the form
\be
S[\hx,\hat y]=\hf\hx\hD^{-1}_0\hx+\hf\hat y\hG^{-1}\hat y-\hx\sigma g\hat y,
\ee
in condensed notation. The elimination of the environment coordinates by means of their equation of motion leads to the effective action, $S_{eff}[\hx]=\hx\hD^{-1}\hx/2$, with $\hD^{-1}=\hD_0^{-1}-\sigma\hat\Sigma\sigma$, where the self energy, $\hat\Sigma=g\hG g$, contains the influence functional,
\be
S_{infl}[\hx]=-\hf\sum_{\sigma\sigma'}\sigma\sigma'\int\frac{d\omega}{2\pi}dte^{-i\omega(t'-t)}\Sigma^{\sigma\sigma'}_\omega x^\sigma(t')x^{\sigma'}(t).
\ee
The spectral function,
\be
\rho(\Omega)=\sum_n\frac{g_n^2}{2m\omega_n}\delta(\omega_n-\Omega),
\ee
can be used to parametrize the model. The self energy assumes the form
\be
\hat\Sigma_\omega=\frac1m\int_0^\infty d\Omega2\Omega\rho(\Omega)\hD_\omega(\Omega),
\ee
where $\hD_\omega(\Omega)$ is given by eq. \eq{doo}, in particular
\be
\Sigma^n_\omega=\frac2mP\int_0^\infty d\Omega\frac{\Omega\rho(\Omega)}{\omega^2-\Omega^2},
\ee
$\Sigma^f_\omega=-i\pi\sign(\omega)\rho(|\omega|)$, and $\Sigma^i_\omega=-\pi\rho(|\omega|)$.

The influence functional is non-local in time but the expansion of the exponent in $t'-t$ and the self energy in $\omega$ renders it local \cite{frict},
\be
L_{infl}=-\hf(x\vec\Sigma^nx^d+x^d\vec\Sigma^nx+x^d\vec\Sigma^fx-x\vec\Sigma^fx^d+x^di\vec\Sigma^ix^d),
\ee
where
\be
\vec\Sigma^{\sigma\sigma'}=\sum_{\ell=0}^\infty\frac{(-1)^\ell}{\ell!}\partial^\ell_{i\omega}\Sigma^{\sigma\sigma'}_0\partial_t^\ell.
\ee

The equation of motion, generated by the variation of $x^d$, at $x^d=0$, is
\be
m\ddot x=-(m\omega^2_0+\vec\Sigma^r)x,
\ee
with $\vec\Sigma^r=\vec\Sigma^n+\vec\Sigma^f$, and contains the following terms up to $\ord{\partial_t^2}$: The harmonic oscillator frequency is renormalized, $\omega^2_0\to\omega^2_0-\Delta\omega^2$, in $\ord{\partial_t^0}$ with
\be
\Delta\omega^2=\frac2{m^2}\int_0^\infty d\Omega\frac{\rho(\Omega)}\Omega,
\ee
realizing the system potential, seen in eq. \eq{holagrc}. A Newtonian friction force, $F=-k\dot x$, arises in $\ord{\partial_t}$ with $k=\pi\rho'(0)$. The $\ord{\partial_t^2}$ mass renormalization, $m\to m+\delta m$, takes place with 
\be
\delta m=\frac4m\int_0^\infty d\Omega\frac{\rho(\Omega)}{\Omega^3},
\ee
reflecting that the dressing by the environment increases the inertia ($\rho\ge0$).

The origin of dissipative forces in this harmonic model can be found by noting that the energy, received by $x$, spreads over the whole system \cite{caus}. This is similar to the spread of energy within an interactive system and leads to dissipation if there are sufficient strength at low frequencies. A more physical insight can be found by viewing irreversibility as a spontaneous symmetry breaking, an adiabatic approximation which becomes exact in the thermodynamical limit: To test reversibility one has to monitor all normal modes. The observation of a normal mode needs time at least in the order of magnitude of its period length. Hence the observations, carried out in an arbitrary long, but finite time leave infinitely many normal modes unresolved and are unable to establish reversibility if there is a condensation point in the spectrum at vanishing frequency. The infinitely many unresolved slow normal mode drive a gradual loss of system energy, diffusion.

\subsection{Classical point charge}
The spectral function of the toy model can freely be chosen and different dissipative features can be modeled. This freedom is due to the lack of local space-time structure. When the degrees of freedom are distributed in the space-time then their spectral weight and the emerging dissipative forces are severely restricted by the space-time symmetries. This motivates our next model, the effective theory of a classical point charge \cite{point}. The electrodynamics of a single charge is defined by the action $S=S_{ch}+S_{EMF}+S_i$, where a point charge, the electromagnetic field (EMF) dynamics and their interactions are described by
\bea
S_{ch}&=&-m_Bc\sum_\sigma\sigma\int ds\sqrt{\dot x^{\sigma2}(s)}\nn\nn
S_{EMF}&=&-\frac1{8\pi c}\sum_\sigma\sigma\int\frac{d^4p}{(2\pi)^4}Q(p^2)p^2A^\sigma_\mu(-p)\left(g^{\mu\nu}-\frac{p^\mu p^\nu}{p^2}\right)A^\sigma_\nu(p),\nn
S_i&=&-\frac{e}c\sum_\sigma\sigma\int ds\dot x^{\sigma\mu}(s)A^\sigma_\mu(x^\sigma(s)),
\eea
$Q(p^2)$ being the regulator for the singular UV regime of the EMF. The influence functional with the real part,
\be\label{eminfl}
\Re S_{infl}=\frac{2\pi e^2}c\int dsds'\dot{\hx}(s)\Re\hat D(x(s)-x(s'))\dot{\hx}(s'),
\ee
can easily be found by eliminating of the EMF. It goes beyond the usual action-at-a-distance action \cite{schwar,tetr,fokk,wheeler} insofar as it contains the radiation field contribution, too. The EMF Green function,
\be
\Re\hat D(x)=\frac{\delta_\ell(x^2)}{4\pi}\begin{pmatrix}-1&-\mr{sign}(x^0)\cr\mr{sign}(x^0)&1\end{pmatrix},
\ee
needs some regularization. The diagonal, near field singularities are removed by smearing the Dirac-delta, $\delta(z)\to\delta_\ell(z)$, over a distance $\ell$. The far field singularity calls for special attention: First, because it arises from the coincidence of the singular points of the two distributions in the off-diagonal blocks. Second, any smearing of $\mr{sign}(x^0)$ breaks Lorentz invariance. We keep $\mr{sign}(x^0)$ unchanged and require $\delta_\ell(0)=0$. Note that we do not need this latter condition in QED, set in the usual formalism, to calculate transition amplitudes between pure states. A simple regularization which satisfies all requirements is $\delta_\ell(z)=\delta(z-\ell^2)$.

The expansion of \eq{eminfl} in $(s-s')/\ell$ is straightforward and produces the non-local influence Lagrangian
\be\label{pcil}
L_{infl}(s)=x^d(s)\frac{4e^2}c\int_{-\infty}^0 du\delta'_\ell(u^2)[x(s+u)-u\dot x(s+u)-x],
\ee
up to terms $\ord{x^2}$. The corresponding equation of motion for $x^d$, at $x^d=0$, is
\be
m_Bc\ddot x^\mu=-\frac{e^2}{c\ell}\ddot x^\mu+\left(g^{\mu\nu}-\dot x^\mu\dot x^\nu\right)\left[K_\nu+\ord{x^3}\right]
\ee
with
\be\label{keq}
K(s)=\frac{2e^2}{3c}\left\{\dddot x(s)-6\int_{-\infty}^0 du\delta'_\ell(u^2)\left[x(s+u)-u\dot x(s+u)-x(s)+\frac{u^2}2\ddot x(s)+\frac{u^3}3\dddot x(s)\right]\right\}.
\ee
The convergence of the integral in the influence Lagrangian, \eq{pcil}, is not uniform when the cutoff is removed, $\ell\to0$, and the first term of \eq{keq}, the well known Abraham-Lorentz force, owes its existence to this irregularity. The emergence of the cutoff-independent Abraham-Lorentz force is reminiscent of quantum anomalies: Both arise as finite cutoff effects, left behind after the removal of the cutoff. In the case of the renormalized classical EMF one finds a light cone anomaly, as well, the modification of the retarded Green function, $D_{r\ell}(x,y)\to D_{r\ell}(x,y)[1+(x-y)^2/s^2_0]$, amounts to an $\ord{e^2/c^2s_0}$ change of the mass of the charge. In other words, the EMF field which is placed slightly off shell by the regulator remains sensitive to the off-shell modifications of its dynamics even after the removal of the cutoff. The second term in \eq{keq}, the integral is $\ord{\ell}$ and generates a crossover, reminiscent of the Landau pole of QED, in the scaling laws at the classical electron radius, $r_0=e^2/mc^2$, the only scale parameter of the model. The usual electromagnetic interaction is recovered in the IR side and an acausal theory emerges in the UV regime of this hypothetical model, without quantum physics.

\subsection{Test particle in an ideal quantum gas}
What changes in the previously discussed models when their quantum version is considered? The classical equation of motion remains an operator equation for harmonic systems, hence the effective equation of motion, derived for the harmonic toy model, remains valid for the expectation values of the coordinate. The toy model has already extensively been studied in the CTP formalism \cite{grabert}. Similarly trivial extension of the classical results to the quantum case is possible for the effective theory of a point charge, too. One has naturally to restrict the motion of the charge to the non-relativistic domain to keep its path integral Gaussian.

Let us now move to a more realistic problem of a quantum test particle, interacting with an ideal gas and determine its effective Lagrangian in the leading order of the Landau-Ginzburg double expansion \cite{frict}. The effective dynamics of a particle which interacts with an ideal gas has already been considered by using the traditional effective action approach in imaginary time \cite{guinea} and by means of the CTP formalism \cite{hedegard}. The quantum master equation \cite{vacchini} has been used extensively, too: The dephasing has been described in the pioneering work \cite{zeh}, followed by the inclusion of relaxation \cite{diosi,hornberger,vacchinie}. The calculation, reviewed here, goes beyond these works insofar as it is based on the CTP formalism, needed to keep track of dissipative forces, and retains all contributions of the CTP self energy and offers a simpler and more flexible alternative to arrive at the effective dynamics. 

The action of this model, $S=S_p+S_g+S_i$, is given by
\bea
S_p[\v{x}]&=&\int dt\left[\frac{M}2\dot{\v{x}}^2(t)-V(\v{x}(t))\right],\nn
S_g[\psid,\psi]&=&\int dtd^3y\psid(t,\v{y})\left[i\hbar\partial_t+\frac{\hbar^2}{2m}\Delta+\mu\right]\psi(t,\v{y}),\nn
S_i[\v{x},\psid,\psi]&=&\int dtd^3yU(\v{y}-\v{x}(t))\psid(t,\v{y})\psi(t,\v{y}),
\eea
where $M$ denotes the mass of the test particle which propagates under the influence of an external potential $V$. The gas is described by the field $\psi$ and its density is coupled to the test particle by the potential $U$. The influence functional of the test particle,
\be\label{inflftpig}
e^{\ih S_{infl}[\hat{\v{x}}]}=\int D[\hpsi]D[\hpsid]e^{\ih\hpsid(\hat F^{-1}+\hat\Gamma[\hat{\v{x}}])\hpsi},
\ee
can be found by integrating over the gas degrees of freedom. The propagator of the gas, the inverse of the quadratic form of $S_g$, is
\be
\hat F(t,\v{y},t',\v{y}')=\int\frac{d\omega}{2\pi}\frac{d^3q}{(2\pi)^3}e^{-i\omega(t-t')+i\v{q}(\v{y}-\v{y}')}\hat F_{\omega\v{q}},
\ee
with 
\be
\hat F_{\omega\v{q}}=\begin{pmatrix}\frac1{\omega-\frac{\hbar\v{q}^2}{2m}+i\epsilon}&0\cr-i2\pi\delta(\omega-\omega_\v{q})&\frac1{\frac{\hbar\v{q}^2}{2m}-\omega+i\epsilon}\end{pmatrix}-\xi n_\v{q}i2\pi\delta\left(\omega-\frac{\hbar\v{q}^2}{2m}\right)\begin{pmatrix}1&1\cr1&1\end{pmatrix},
\ee
where $\xi=1$ for bosons and $\xi=-1$ for fermions and $n_\v{q}$ stands for the occupation number. The temperature should be high enough for bosons to suppress the condensate. The test particle-gas interaction is represented by the term $\hpsid\hat\Gamma[\hat{\v{x}}]\hpsi=\sum_\sigma\sigma S_i[\v{x}^\sigma,\psi^{\sigma\dagger},\psi^\sigma]$. We assume that the potential localizes the particle strong enough to justify the expansion in the coordinate and seek the leading, $\ord{\v{x}^2}$ contributions to the influence functional. Hence we retain the $\ord{\hat\Gamma^2}$ part,
\be\label{idgifl}
S_{infl}[\hat{\v{x}}]=-\hf\hj\hat\sigma\hG\hat\sigma\hj,
\ee
only. The tadpole contribution is canceled by introducing a homogeneous, neutralizing classical background charge and  
\be
G^{\sigma_1\sigma_2}(x_1,x_2)=i\xi\hbar n_s\hat F^{\sigma_1\sigma_2}(x_1,x_2)\hat F^{\sigma_2\sigma_1}(x_2,x_1)
\ee
denotes the density two-point function and $j^\sigma(t,\v{y})=U(\v{y}-\v{x}^\sigma(t))$. The two-point function is given by well known one-loop integrals,  $G^n_{\omega\v{q}}=G^+_{\omega\v{q}}+G^+_{-\omega\v{q}}$, $G^f_{\omega\v{q}}=G^-_{\omega\v{q}}-G^-_{-\omega\v{q}}$ and $iG^i_{\omega\v{q}}=G^-_{\omega\v{q}}+G^-_{-\omega\v{q}}$, with 
\bea\label{gpm}
G^+_{\omega\v{q}}&=&-\xi\frac{n_s}{\hbar^2}P\int\frac{d^3q}{(2\pi)^3}\frac{n_\v{q}}{\omega-\omega_{\v{q}+\v{q}}+\omega_\v{q}},\nn
G^-_{\omega\v{q}}&=&-i\xi\pi\frac{n_s}{\hbar^2}\int\frac{d^3q}{(2\pi)^3}n_\v{q}(n_{\v{q}+\v{q}}+\xi)\delta(\omega-\omega_{\v{q}+\v{q}}+\omega_\v{q}),
\eea
where $n_s=2s+1$ and $s$ denotes the spin of the particles of the gas. These integrals are analytic functions of the dimensionless variables $x=\omega/|\v{q}|v_F$, $y=|\v{q}|/k_F$ where $v_F=\hbar k_F/m$, $k_F=\sqrt[3]{6\pi^2n/n_s}$, $n$ being the density of the gas, if the gas is in thermal equilibrium at finite temperature. The characteristic frequency of the particle-gas interaction is $|\v{q}\dot{\v{x}}|$ hence the second expansion of the Landau-Ginzburg scheme, the expansion in the time derivative, is possible for $|x|\ll1$, if the test particle moves slower than the gas particles, $|\dv{x}|\ll v_F$. 

In calculating the influence Lagrangian we follow the strategy of Section \ref{toymods} and start with the influence functional
\be\label{inflact}
S_{infl}=-\hf\sum_{\sigma\sigma'}\sigma\sigma'\int\frac{d\omega}{2\pi}\frac{d^3q}{(2\pi)^3}dtdt'U_\v{q}^2e^{-i\omega(t'-t)+i\v{q}(\v{x}^\sigma(t')-\v{x}^{\sigma'}(t))}G^{\sigma,\sigma'}_{\omega\v{q}},
\ee
whose $\ord{\v{x}^2}$ part gives the influence Lagrangian,
\be\label{inflbe}
L_{infl}=\frac14\sum_{\sigma\sigma'}\sigma\sigma'\int_\v{q}U_\v{q}^2[\v{q}(\v{x}^\sigma-\v{x}^{\sigma'}+\Delta\v{x}^\sigma)]^2G^{\sigma\sigma'}_{0\v{q}},
\ee
with 
\be
\Delta\v{x}=\sum_{n=1}^\infty\frac{\v{x}^{(n)}}{n!}\partial_{i\omega}^n,
\ee
and $\v{x}=\v{x}(t)$, $\v{x}^{(n)}=d^n\v{x}/dt^n$. The influence Lagrangian can be brought into the form
\bea
L_{infl}&=&\frac1{24\pi^2}\int_0^\infty dqq^2U_q^2[2\Delta\v{x}\Delta\v{x}^dG^n_{0q}+(2\Delta\v{x}\Delta\v{x}^d+4\v{x}^d\Delta\v{x})G^f_{0q}\nn
&&-2(\v{x}^{d2}+\v{x}^d\Delta\v{x}^d)iG^i_{0q}],
\eea
by the help of the block structure \eq{blockg}. The leading order terms in the expansion of the time derivative lead to
\be\label{linflk}
L_{infl}=-k\v{x}^d\dv{x}+\delta M\dv{x}^d\dv{x}+id_0\v{x}^{d2}-id_2\dv{x}^{d2}+\ord{\partial_t^3}+\ord{\v{x}^4},
\ee
with the coefficients
\bea\label{emcoeff}
k&=&-\frac1{6\pi^2v_F}\int_0^\infty dqq^3U_q^2\partial_{ix}G^f(x,y)_{|x=0},\nn
\delta M&=&\frac1{12\pi^2v_F^2}\int_0^\infty dqq^2U_q^2\partial^2_{ix}G^n(x,y)_{|x=0},\nn
d_0&=&-\frac1{12\pi^2}\int_0^\infty dqq^4U_q^2G^i(x,y)_{|x=0},\nn
d_2&=&\frac1{24\pi^2v_F^2}\int_0^\infty dqq^2U_q^2\partial_{ix}^2G^i(x,y)_{|x=0}.
\eea
The expectation value of the real part of the Euler-Lagrange equation for $\v{x}^d$ at $\la\v{x}^d\ra=0$,
\be\label{relpingem}
M_R\la\ddv{x}\ra=-\la\v{\nabla}V(\v{x})\ra-k\la\dv{x}\ra+\ord{\partial_t^3}+\ord{\la\v{x}^3\ra},
\ee
comes from $\Re L_{infl}$ and includes a mass renormalization, $M_R=M+\delta M$, and a friction force. The imaginary part of $L_{infl}$ generates decoherence, a suppression factor, $\exp-\Im S_{infl}$, in the path integral. The terms involving $\v{x}^{d2}$ and $\dv{x}^{d2}$ represent the decoherence strength in the coordinate and in the momentum basis, respectively.

Note that the Newtonian friction force allows us to determine the velocity with respect to the gas. This is possible due to the breakdown of the boost invariance for the test particle by its environment. In our calculation of the effective dynamics of a point charge the EMF started in a Lorentz invariant intial state, $A_\mu=0$, and the dissipative effective interaction, the Abraham-Lorentz force, must contain higher, odd-order time derivative.

\subsection{Quantum ideal gas}
The emergence of dissipative forces and decoherence when a particle interacts with a gas raises the question whether dissipative forces could be found within the ideal gas itself, without any test particle \cite{hydro}. 

The simplest, hand-waving way to argue about the possibility of finding a non-trivial effective dynamics for a non-linear combination of the fundamental coordinates in a harmonic system is to recall that a non-linear coordinate transformation generates inertial forces. For instance, if one observe the function $y=x^2/2$ of the coordinate of a harmonic oscillator, defined by the Lagrangian $L=m\dot x^2/2-m\omega^2x^2/2$ then the effective dynamics is given by the Lagrangian $L=m\dot y^2/4y-m\omega^2y$ which seems to generate a non-trivial dynamics. The inertial forces  superficially seem as complicated as the forces, arising from genuine interactions. 

To find a more systematical approach to the effective forces in a quantum gas we exploit the stability of the particles within the gas. Owing to the conserved particle number the elementary excitations, driven by some coupling to the environment, are handled by a bi-local operator, $\Phi(x,y)=\psi(x)\psi^\dagger(y)$, where the elementary field, $\psi(x)$, belongs to the gas particles. By borrowing from the idea of operator product expansion we can replace the non-local operator by infinitely many local composite operators, 
\be
\Phi^{\mu_1,\ldots,\mu_n}_n(x)=\frac{\partial^n}{\partial z^{\mu_1}\cdots\partial z^{\mu_n}}\psi(x+z)\psi^\dagger(x-z).
\ee
These operators serve as possible terms in describing the interactions with the environment and as possible observables to use in diagnosing the gas dynamics. When they appear in the gas-environment coupling then the energy-momentum, received from the environment, is spread over infinitely many normal modes. If they are used as observables then they receive contributions from equally infinitely many normal modes. 

The simplicity of the ideal gas dynamics appears only if the elementary fields, $\psi$ and $\psid$, are used to diagnose the system. In fact, the Wick theorem, stating the factorization of  the $n$-point functions of $\psi$ and $\psid$ into the product of the free propagators, implies the absence of  1PI vertex functions. But there are non-factorizable, 1PI composite operator $n$-point functions for arbitrary $n$! When can we use the elementary fields, $\psi$ or $\psid$, and uncover the non-interacting particle picture and when are we forced to deal with the bi-local or composite operators? The answer is easy to find if the symmetry, behind the conservation of the gas particle number, is gauged, as in the case for the electron gas, whatever weak the gauge coupling constant might be. The dynamics of a single quasi-particle of such a model is given in terms of non-local observables, such as the scattering amplitudes. The local observables are constructed by the help of the composite operator family, $\Phi_n$. Thus the price of using local observables is the loss of the simplicity, offered by the quasi-particle picture. When is the locality a more important feature than the simplicity in describing the interactive system as a gas of quasi-particles with weak residual interactions? When we want to study the structure of a single quasi-particle, rather than the long distance correlations. The local fields which appear in the microscopic, local interaction Hamiltonian are the natural variables to understand the internal structure of the quasi-particles, c.f. the restricted hydrodynamical regime of the ideal gas collective modes in an interacting gas, given by eq. \eq{ihreg} below.

It is natural to inquire about the effective dynamics of a restricted subset of the composite operators. We follow up this question by considering the effective dynamics of the $n=0$, ultra-local operators in a non-interacting electron gas. There are 16 possibilities and we retain 4, $j^\mu=\bar\psi\gamma^\mu\psi$, the current. Note that the current has an environment, provided by the remaining local composite operators, $\Phi_n$, even if the Fock-space is not a system$\otimes$environment type direct product \cite{lajos}.

The generator functional of the current Green functions is
\be
e^{\ih W[\ha]}=\int D[\hpsi]D[\hsib]e^{\ih\hsib(\hat F^{-1}+\ha\bre)\hpsi},
\ee
where $\hsib\ha\bre\hpsi=\sum_\sigma a^\sigma_\mu\bar\psi^\sigma\gamma^\mu\psi^\sigma$. The Gaussian integral is easy to perform,
\be
W[\ha]=-i\hbar\Tr[\hat F^{-1}+\ha\bre],
\ee
and indeed we find 1PI Green functions at arbitrary orders. However there is an important difference between the ideal gas dynamics and a truly interactive system: The former has one-loop 1PI graphs only and the vertex functions receive contributions from arbitrarily  high orders of the loop-expansion in the latter case. Nevertheless the functional $W[\ha]$ of the ideal gas, together with its Legendre transform present a formidable problem, they contain arbitrary high order contributions, without adjustable coupling constant to organize an expansion. The correlations, generated by the composite operators, used in diagnosing the gas, generate strong long range correlations. Fortunately, one can consider the weak external perturbations limit where one finds
\be
W[\ha]=-\hf\ha\hG\ha,
\ee
by ignoring the $\ord{\ha^3}$ contributions, with
\be\label{ccgfct}
G^{\sigma_1\sigma_2}_{\mu_1\mu_2}(x_1,x_2)=-i\hbar\tr[\gamma^{\mu_1}\hat F^{\sigma_1\sigma_2}(x_1,x_2)\gamma^{\mu_2}\hat F^{\sigma_2\sigma_1}(x_2,x_1)],
\ee
denoting the current two-point function. The corresponding effective action,
\be\label{currea}
\Gamma[\hJ]=\hf\hJ\hG^{-1}\hJ,
\ee
can be written as
\be
\Gamma[J,J^d]=\hf J^dK^rJ+\hf JK^aJ^d-\frac{i}2J^dK^rK^aJ^d,
\ee
where $J=(J^++J^-)/2$, $J^d=J^+-J^-$, $\hat K=\hat\sigma\hG^{-1}\hat\sigma$ and the linearized equation of motion for a physically realizable external source, $\ha=(a,-a)$,
\be\label{eomc}
a=-K^rJ,
\ee
sometime called as the resistivity formula \cite{toda}, is the inverse of Kubo's linear response equation, $J=-G^ra$.

The Green function, \eq{ccgfct}, can be calculated in a straightforward manner for zero temperature fermion gas and leads to the equation of motion for the inhomogeneous part of the current, $j^\mu=(n_0+n,\v{j})$,
\bea\label{eomgradexp}
-\frac{v_F}{\pi^2\hbar}\phi&=&\left[\sum_{jk=0}^\infty b_{\ell,j,k}\left(\frac{i\omega}{v_F|\v{q}|}\right)^j\left(\frac{\v{q}^2}{k_F^2}\right)^k\right]n\\
\frac{T^2}{v_F\hbar}\v{a}-\frac{b_{t,2,0}}{b_{\ell,2,0}}\frac{v_F}{\pi^2\hbar}\frac{\omega\v{q}}{\v{q}^2}\phi&=&\biggl[\sum_{jk=0}^\infty b_{t,j,k}\left(\frac{i\omega}{v_F|\v{q}|}\right)^j\left(\frac{\v{q}^2}{k_F^2}\right)^k+\frac{\v{q}\otimes\v{q}}{\v{q}^2}\sum_{jk=0}^\infty b_{j,k}\left(\frac{i\omega}{v_F|\v{q}|}\right)^j\left(\frac{\v{q}^2}{k_F^2}\right)^k\biggr]\v{j},\nonumber
\eea
where $a^\mu=(\phi,\v{a})$, $k_F$ denotes the Fermi wave number, $v_F=\hbar k_F/m$, and $b_{\ell,j,k}$, $b_{t,j,k}$ and $b_{j,k}$ are dimensionless constants. This equation holds for 
\be
q^\mu=(\omega,\v{q})\in{\cal D}^0_{hydr}=\left\{(\omega,\v{q})|\left|\frac{|\omega|}{v_F|\v{q}|}-\frac{|\v{q}|}{k_F}\right|<1\right\}.
\ee

It is instructive to compare eq. \eq{eomgradexp} with the Navier-Stokes equation of phenomenological hydrodynamics. Despite the obvious similarities, both mix the odd and the even powers of $\omega$ and describe dissipative dynamics, there are important differences:
\begin{enumerate}
\item Hydrodynamics refers to the energy-momentum current, rather than the Noether current, appearing in the effective action \eq{currea}. 
\item The hydrodynamical equations are closed by relying on local equilibrium and thermodynamic considerations while the assumption of the stability of the ground state, the smallness of the inhomogeneity of the current, is enough to arrive at the closed equation \eq{eomc}. 
\item The gradient expansion of the phenomenological hydrodynamics assumes that the equations are analytic at $q^\mu=0$ and the IR dynamics in the hydrodynamical regime is independent of the way the point $q^\mu=0$ is approached. The Fermi-level makes the $\omega$-dependence of the the Green function to appear through the combination $\omega/|\v{q}|$ thereby making the IR point, $q^\mu=0$, singular and the IR dynamics dependent on the slope of the curve at this point on the $(\omega,|\v{q}|)$ plane. The result is a Laurent series in $|\v{q}|$. The factor $1/|\v{q}|$, accompanying each $i\omega$ in the equation of motion, can not be foreseen in the phenomenological approach. 
\item The $\ord{\partial_t}$ phenomenological hydrodynamical equations describe collective modes, sound waves. The effective equation of motion, \eq{eomgradexp}, describes sound waves, as well, but it has to be truncated at least at $\ord{\partial_t^2}$ to yield a qualitatively stable, truncation independent form. This composite sound, neither the known zero nor first sound, is damped, displays long range correlations and dominates the current decoherence. The long range correlations arise because the effective equation of motion, written in the space-time, contains the powers of $\Delta$ and the convolutions of the Fourier transforms of $\omega/|\v{q}|$, the latter being divergent at large distance beyond the order $\ord{\omega^2}$.
\end{enumerate}

The physics of the ideal gas hydrodynamics, described by eqs. \eq{eomgradexp} seems to be rather different from the usual Fermi-liquid picture. To resolve this conflict we return to a remark, made previously, namely that interactions lead to higher order 1PI graphs. One of their effects, the finite life-time of the quasi-particles, restores the analycity at $q^\mu=0$. The collisions in a realistic, interacting gas form the usual collective modes and quasi-particles at the distance scale of the mean-free path, $r_{mfp}>1/k_F$. The ideal gas description with its composite sound is thereby limited to the regime
\be\label{ihreg}
{\cal D}^{int}_{hydr}=\left\{(\omega,\v{q})|\left|\frac{|\omega|}{v_F|\v{q}|}-\frac{|\v{q}|}{k_F}\right|<1,\frac{|\v{q}|}{k_F}>\frac1{k_Fr_{mfp}}\right\}.
\ee
Therefore the composite sound is relevant at distances between $1/k_F$ and $r_{mfp}$ and plays an important role in forming the structure of the quasi-particles, in agreement with the remark, made above. It is interesting to look at the QCD vacuum from this point of view, where the single gluon modes play the role of composite sound within the quasi-particles of the vacuum, glueballs, and display strong correlation beyond the quasi-particle size.

\end{document}